\documentclass[letterpaper, 10 pt, conference]{ieeeconf}
\IEEEoverridecommandlockouts
\let\labelindent\relax
\usepackage{amsmath,amssymb,amsthm,mathtools}
\usepackage{cite}
\usepackage{enumitem}
\usepackage{tikz}
\usetikzlibrary{arrows.meta,positioning,calc,shapes,arrows,positioning,calc,trees,backgrounds,intersections,fit}
\usepackage{graphicx}
\usepackage{float}
\usepackage{subcaption}
\usetikzlibrary{positioning,arrows.meta}
\makeatletter\let\NAT@parse\undefined\makeatother
\usepackage[colorlinks=true,linkcolor=blue,citecolor=blue]{hyperref}

\usepackage[capitalize,nameinlink]{cleveref}
\crefname{example}{Example}{Examples}
\crefformat{problem}{#2Problem#3}
\Crefformat{problem}{#2Problem#3}
\usepackage[normalem]{ulem}

\newcommand{\PartI}{%
  \begingroup
  \renewcommand{\citepunct}{, }%
  \renewcommand{\citedash}{--}%
  \cite{YoonPartIACC,YoonPartIext}%
  \endgroup
}
\newtheorem{theorem}{Theorem}

\newtheorem{property}{Property}

\theoremstyle{definition}
\newtheorem{definition}{Definition}
\newtheorem{assumption}{Assumption}
\newtheorem*{problem}{Problem}
\theoremstyle{definition}

\theoremstyle{remark}

\crefname{problem}{Problem}{Problems}
\crefname{procedure}{Procedure}{Procedures}
\crefname{assumption}{Assumption}{Assumptions}
\crefname{property}{Property}{Properties}
\Crefname{property}{Property}{Properties}

\newcommand{\R}{\mathbb{R}}
\newcommand{\xb}{\bar{x}}

\newcommand{\Ab}{\bar{A}}
\newcommand{\By}{B_y}
\newcommand{\Clim}{C_{\lim}}
\newcommand{\Cz}{C_z}

\newcommand{\Kfull}{K_{\mathrm{full}}}
\newcommand{\Hu}{H_u}
\newcommand{\ycmd}{y_{\mathrm{cmd}}}
\newcommand{\yg}{y_g}
\newcommand{\ag}{\alpha_g}
\newcommand{\zmax}{z_{\max}}
\newcommand{\zmin}{z_{\min}}
\newcommand{\umax}{u_{\max}}
\newcommand{\umin}{u_{\min}}
\newcommand{\Z}{\mathcal{Z}}
\newcommand{\U}{\mathcal{U}}
\newcommand{\W}{\mathcal{W}}
\newcommand{\Sset}{S}
\newcommand{\SE}{\Sset_{E}}

\newcommand{\Oh}{\mathcal{O}}

\newcommand{\Creg}{C_{\mathrm{reg}}}
\newcommand{\Dreg}{D_{\mathrm{reg}}}
\newcommand{\Bz}{B_\zeta}

\newcommand{\etat}{\tilde{\eta}}

\newcommand{\Bd}{B_d}
\newcommand{\Tz}{T_{\xi}}

\newcommand{\ThB}{\bar{\Theta}}
\newcommand{\ThH}{\hat{\Theta}}
\newcommand{\nub}{\bar{\nu}}

\newcommand{\ones}{\mathbf{1}}
\newcommand{\T}{^{\top}}

\newcommand{\Pieta}{\Pi_{\eta}}

\DeclareMathOperator*{\argmin}{arg\,min}

\usepackage{algorithm}
\usepackage{algorithmic}
\usepackage[most]{tcolorbox}
\usepackage{colortbl}
\graphicspath{{figs/}}
\newif\ifcolorboxes
\colorboxestrue
\newif\ifshownotes
\shownotesfalse

\ifcolorboxes\colorlet{keyrow}{black!8}\else\colorlet{keyrow}{white}\fi
\newcommand{\stmtbox}[2]{\ifcolorboxes\tcolorboxenvironment{#1}{enhanced jigsaw,
breakable, frame hidden, boxrule=0pt, sharp corners, colback=#2,
left=3pt, right=3pt, top=3pt, bottom=3pt, boxsep=0pt,
grow to left by=0pt, grow to right by=0pt,
before skip=8pt, after skip=8pt}\fi}
\stmtbox{definition}{blue!7}
\stmtbox{assumption}{teal!10}
\stmtbox{lemma}{yellow!13}
\stmtbox{proposition}{green!9}
\stmtbox{property}{green!9}
\stmtbox{theorem}{red!8}
\stmtbox{problem}{orange!10}
\stmtbox{remark}{black!5}
\stmtbox{procedure}{cyan!8}
\providecommand{\Mg}{\alpha_g}

\providecommand{\Pieta}{\Pi_{\eta}}

\begin{document}

\title{\LARGE \bf Feasibility of Simultaneous Input-Output Constraints for Tracking in a Class of LTI Systems: Part~II}

\author{Junhyeok~Yoon$^{1}$, Alan~B.~Cao$^{1}$, Heather~Hussain$^{2}$, and Anuradha~M.~Annaswamy$^{1}$%
\thanks{$^{1}$J. Yoon, A. B. Cao, and A. M. Annaswamy are with the Department of Mechanical Engineering, Massachusetts Institute of Technology, Cambridge, MA 02139, USA (e-mail: yjunh01@mit.edu; alancao@mit.edu; aanna@mit.edu).}%
\thanks{$^{2}$H. Hussain is with The Boeing Company, Engineering \& Technology Innovation, USA (e-mail: heather.s.hussain@boeing.com).}%
\thanks{This work was supported by the Boeing Strategic University Initiative.}}

\maketitle
% \raggedbottom 

%===========================================%
%                 ABSTRACT                  %
%===========================================%

\begin{abstract}
This paper illustrates the properties of a new CBF-governor that enables simultaneous input and output constraint satisfaction for a class of multi-input linear time-invariant systems with state feedback and integral action. This governor is designed so as to modify the reference command while preserving the nominal feedback controller. Necessary and sufficient conditions are derived in a companion paper under which the governor is shown to be feasible at every state of a prescribed operating set and ensures simultaneous input and output constraint satisfaction, forward invariance, and bounded closed-loop solutions. These  theoretical results are illustrated in this paper through several numerical examples. In each of these examples, we show how the goals of the CBF-governor are met and corroborate the corresponding necessary and sufficient condition. The computational burden associated with the proposed CBF-governor is also articulated, with its online component comparable to either that of a QP solver or determined using an exact closed-form solution. The offline component is related to the checking of the SIOCF condition which is of $O(N)$.
\end{abstract}
% keywords are entered in PaperPlaza and omitted here for the page limit

%===========================================%
%               INTRODUCTION                %
%===========================================%

\section{Introduction}\label{sec:intro}
Control barrier functions (CBFs) provide a systematic framework for enforcing safety constraints in dynamical systems by defining a safe set through a barrier function $h(x)$ and imposing conditions on its evolution to ensure forward invariance \cite{AmesCBF19,XiaoBelta22}. A vast literature has developed along this vein, extending CBF methods to a broad range of safety-critical control problems. Safety concerns are often modeled as constraints on key outputs of the dynamic system. 

Simultaneous satisfaction of both input and output constraints has received comparatively less attention \cite{ChenJankovicSantilloAmes21,AgrawalPanagou21,ChoiLeeSreenathTomlinHerbert21,XiaoBeltaCassandras22,IsalyGhanbarpourSanfeliceDixon24,BreedenPanagou23,MousaviTanAmes25,MousaviAmes26,
lavretsky2025servo,FisherAnnaswamy25}. In most of these methods, forward invariance is established either using only sufficient conditions or requires a high computational burden. 
The authors of \cite{lavretsky2025servo} consider a general LTI system, include integral action, consider both input and output constraints, and propose a specific augmentation-based control architecture for stability and constraint satisfaction. The feasibility of the latter, however, is not established. 
In a companion paper, we have proposed a new approach for simultaneous satisfaction of both input and output constraints for command tracking in LTI systems with integral action and state feedback \PartI. This approach employs a CBF-based governor and consists of necessary and sufficient conditions for I-O constraint satisfaction, closed-loop boundedness, and command tracking as closely as possible.
The results of \PartI\   also show how free design parameters of the proposed I-O CBF Governor can be chosen so that the necessary and sufficient conditions can always be satisfied by the underlying closed-loop system with forward invariance of a given set, which corresponds to a relaxation of the input constraints. An efficient design of the CBF governor is also described, so as to ensure that this relaxation is near-minimal. 

In this paper, we carry out a detailed description of how the I-O CBF Governor is designed and implemented and delineate a number of examples and simulation studies that demonstrate the usefulness of the necessary and sufficient conditions. In particular, we show how offline computations can be carried out to check if these conditions are satisfied. Counter-examples are provided to show that simultaneous I-O constraints can be fundamentally incompatible. Finally and most importantly, a constructive design procedure is provided to recover feasibility  in these cases by suitably relaxing the input constraints. 
The problem as well as the I-O CBF Governor, and the key theoretical results are described in \cref{sec:setting}. Six numerical examples are provided in \cref{sec:example} to illustrate the theoretical results.

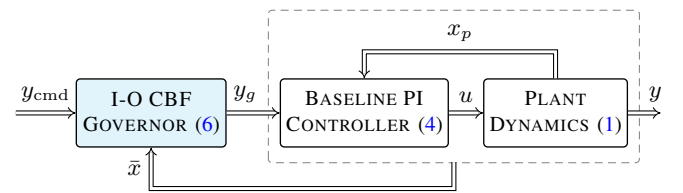
\begin{figure}[b]
\vspace{-0.1in}
\centering
\begin{tikzpicture}[
  font=\fontsize{8}{9.6}\selectfont,
  block/.style={draw, rounded corners=1.5pt, line width=0.5pt, minimum height=0.9cm, align=center, inner sep=2pt},
  vec/.style={double, double distance=1.1pt, line width=0.4pt, -{Implies}},
  lbl/.style={font=\fontsize{9}{10.8}\selectfont, inner xsep=1.5pt, inner ysep=4pt}]
\node[block, minimum width=1.98cm, fill=cyan!10] (gov) at (1.790,0) {\textsc{I-O CBF}\\\textsc{Governor}~\eqref{eq:qpproj}};
\node[block, minimum width=2.22cm] (pi) at (4.610,0) {\textsc{Baseline PI}\\\textsc{Controller}~\eqref{eq:pi}};
\node[block, minimum width=1.9cm] (plant) at (7.140,0) {\textsc{Plant}\\\textsc{Dynamics}~\eqref{eq:plant}};
\draw[gray, line width=0.35pt, densely dashed, rounded corners=0.5mm]
  ([xshift=-1.5mm, yshift=-2mm]pi.south west) rectangle ([xshift=1.5mm, yshift=9mm]plant.north east);
\draw[vec] (0,0) -- node[lbl, above] {$y_{\mathrm{cmd}}$} (gov.west);
\draw[vec] (gov.east) -- node[lbl, above, pos=0.35] {$y_g$} (pi.west);
\draw[vec] (pi.east) -- node[lbl, above] {$u$} (plant.west);
\draw[vec] (plant.east) -- node[lbl, above, pos=0.84] {$y$} (8.53,0);
\draw[vec] (plant.north) -- ++(0,0.35) -| node[lbl, above, pos=0.25] {$x_p$} (pi.north);
\draw[vec] (5.795,-0.65) -- (5.795,-1.0) -| node[lbl, left=2pt, pos=0.75] {$\bar x$} (gov.south);
\end{tikzpicture}
\caption{I-O CBF Governor architecture. The I-O CBF Governor
modifies the desired command $y_{\mathrm{cmd}}$ to produce $y_g$. The augmented
closed-loop state $\bar{x}$ is fed back to the governor.}
\label{fig:blockdiagram}
\end{figure}

%===========================================%
%             PROBLEM SETTING               %
%===========================================%

\section{The Problem and the I-O CBF Governor}\label{sec:setting}

We consider a class of LTI plants with integral action:
\begin{align}
\dot x_p&=A_px_p+B_pu,\notag\\
y&=\Creg x_p+\Dreg u,\label{eq:plant}\\
\dot e_I&=y-\yg ,\notag
\end{align}
where $x_p\in\R^{n_p}$ is the plant state, $u\in\R^m$ is the control input, $y\in\R^m$ is the output required to track a desired signal $\ycmd\in\R^m$, and $e_I\in\R^m$ is the integrator state.
$\yg$ in \eqref{eq:plant} represents a modified command (see \autoref{fig:blockdiagram}) to be chosen so that constraints on the input $u$ and relevant outputs of the system $z$ are satisfied at all times.
The latter is specified as
\begin{subequations}\label{eq:req}
\begin{alignat}{2}
u(t)&\in\U:=\{u\in\R^m:\umin\le u\le\umax\}&\ \ &\forall t\ge t_0,\label{eq:requ}\\
z(t)&\in\Z:=\{z\in\R^m:\zmin\le z\le\zmax\}&\ \ &\forall t\ge t_0,\label{eq:reqz}
\end{alignat}
\end{subequations}
where
\begin{equation}\label{eq:zdef}
z=\Clim x_p ,
\end{equation}
with $\Clim\in\R^{m\times n_p}$.
In \eqref{eq:req}, $\zmin,\zmax,\umin,\umax\in\R^m$, and the inequalities are imposed component-wise, where $\umin<0<\umax$ and $\zmin<0<\zmax$.
The overall objective of this paper is that all states remain bounded, $y$ tracks $\ycmd$ as closely as possible, and that constraints on $u$ and $z$ be satisfied at all times.
In order to achieve the first objective, we propose a state feedback of the form
\begin{equation}\label{eq:pi}
u=-K_Px_p-K_Ie_I ,
\end{equation}
where $K_P\in\R^{m\times n_p}$ and $K_I\in\R^{m\times m}$.
In order to achieve the latter two objectives, we propose a CBF-based governor, which is the focus of this paper.

We rewrite the closed-loop system using \eqref{eq:plant} and \eqref{eq:pi} as
\begin{gather}
\dot\xb=\Ab\xb+\By\,\yg ,\qquad
\xb:=\begin{bmatrix}x_p\\ e_I\end{bmatrix}\in\R^{n_p+m},\label{eq:loop}\\
\Ab:=\begin{bmatrix}A_p{-}B_pK_P & {-}B_pK_I\\ \Creg{-}\Dreg K_P & {-}\Dreg K_I\end{bmatrix},
\quad
\By:=\begin{bmatrix}0_{n_p\times m}\\ -I_m\end{bmatrix}.\notag
\end{gather}

The focus of the paper is to find a modified command $\yg$ such that, given the closed-loop system \eqref{eq:loop} and the constraint requirement \eqref{eq:req},  every solution of \eqref{eq:loop} remains bounded and \eqref{eq:req} is satisfied for all bounded piecewise continuous $\ycmd$.

The following assumptions are made regarding the underlying LTI system, where $\Hu:=\Clim B_p$.

\begin{assumption}\label{ass:standing}\leavevmode
\begin{enumerate}[label=(\textnormal{A\arabic*}),leftmargin=2.6em,topsep=2pt]
\item\label{ass:A1} $\Hu$ is nonsingular;
\item\label{ass:A2} the invariant zeros of $(A_p,\!B_p,\!\Clim,\!0_{m\times m})$ lie in $\mathbb C^{-}$;
\item\label{ass:A3} $\Ab$ is Hurwitz.
\end{enumerate}
\end{assumption}

Assumptions \textnormal{(A1)} and \textnormal{(A2)} pertain only to the output $z$ and impose no restriction on the tracked output $y$.
Assumption \textnormal{(A3)} can be easily satisfied by the choice of $(K_P,K_I)$ whenever \eqref{eq:plant} is stabilizable.

%===========================================%
%             I-O CBF GOVERNOR              %
%===========================================%

\subsection{I-O CBF Governor}\label{sec:governor}

The CBF Governor, denoted as the I-O (Input and Output) CBF Governor, will generate 
\begin{equation}\label{eq:qpproj}
\yg(\xb,\ycmd)=\argmin_{\zeta\in F(\xb)}\ \lVert\zeta-\ycmd\rVert^{2}.
\end{equation}
where the pointwise feasible set \cite{BoydV04} is
\begin{equation}\label{eq:qpF}
F(\xb):=\big\{\zeta\in\R^m:\ A_\zeta\,\zeta\le b(\xb)\big\} ,
\end{equation}
with $b(\xb):=\Bz\xb+b_0$ and the constant matrices
\begin{equation}\label{eq:qpdata}
\begin{gathered}
A_\zeta:=\begin{bmatrix}\Hu K_I\\ -\Hu K_I\\ K_I\\ -K_I\\
I_m\\ -I_m\end{bmatrix},
\qquad
b_0:=\begin{bmatrix}
\alpha_1\alpha_2\,\zmax\\ -\alpha_1\alpha_2\,\zmin\\
\alpha_u\,\umax\\ -\alpha_u\,\umin\\
\Mg\ones_m\\ \Mg\ones_m
\end{bmatrix},\\[6pt]
\Bz:=\begin{bmatrix}
-\big(\Cz\Ab^{2}+(\alpha_1{+}\alpha_2)\Cz\Ab+\alpha_1\alpha_2\Cz\big)\\
+\big(\Cz\Ab^{2}+(\alpha_1{+}\alpha_2)\Cz\Ab+\alpha_1\alpha_2\Cz\big)\\
\Kfull(\Ab+\alpha_uI)\\ -\Kfull(\Ab+\alpha_uI)\\ 0\\ 0
\end{bmatrix},
\end{gathered}
\end{equation}
where $\alpha_1,\alpha_2,\alpha_u>0$ are defined in the high-order CBFs (HOCBFs)
\begin{align*}
\psi_1^{\pm}&:=\dot h_1^{\pm}+\alpha_1h_1^{\pm}, & \psi_2^{\pm}&:=\dot\psi_1^{\pm}+\alpha_2\psi_1^{\pm},\\
\psi_u^{\pm}&:=\dot h_u^{\pm}+\alpha_uh_u^{\pm},
\end{align*}
with CBF pairs
\begin{align*}
h_1^{+}(\xb)&:=\zmax-\Cz\xb, & h_1^{-}(\xb)&:=\Cz\xb-\zmin,\\
h_u^{+}(\xb)&:=\umax+\Kfull\xb, & h_u^{-}(\xb)&:=-\Kfull\xb-\umin ,
\end{align*}
where $\Cz:=[\Clim\ \ 0_{m\times m}]$ and $\Kfull:=[K_P\ \ K_I]$.

We denote $F(\xb(t))\neq\varnothing$, $\forall t\ge t_0$, as the \emph{feasibility condition}. It is clear that if the feasibility condition is satisfied, then all I-O constraints \eqref{eq:req} are satisfied, and the solution $\yg$ is continuous and piecewise affine.

We introduce two sets, $\Omega$ and $\Sset$, and a definition.
The set $\Omega$ is defined as
\begin{equation}\label{eq:omega}
\Omega:=\{\xb\in\R^{n_p+m}:\ h_1^{\pm}(\xb)\ge0,\ \psi_1^{\pm}(\xb)\ge0,\ h_u^{\pm}(\xb)\ge0\},
\end{equation}
which will be used to establish forward invariance of the closed-loop system \eqref{eq:loop}, i.e.\ if $\xb(t_0)\in\Omega$, then $\xb(t)\in\Omega$ $\forall t\ge t_0$.
The design parameter $\alpha^{1}\in\R_{++}^{4}$ of the I-O CBF Governor, where $\R_{++}$ denotes the positive reals, is given by
\begin{equation}\label{eq:alpha1}
\alpha^{1}:=[\,\alpha_1\ \ \alpha_2\ \ \alpha_u\ \ \Mg\,]\T .
\end{equation}
We now introduce a definition, SIOCF, that formalizes the property of feasibility:

\begin{definition}\label{def:siocf}
Let $\Sset\subseteq\Omega$ be nonempty and closed.
The I-O CBF Governor \eqref{eq:qpproj} \textbf{\emph{satisfies simultaneous input and output constraint feasibility (SIOCF)}} on $\Sset$ if, for all bounded piecewise continuous $\ycmd$ and $\forall\xb(t_0)\in\Sset$, the solutions of \eqref{eq:loop} satisfy the following:
\begin{enumerate}[label=(P\arabic*),leftmargin=2.6em]
\item existence and uniqueness of $\yg$ on $[t_0,\infty)$;
\item $\xb(t)\in\Sset$ $\ \forall t\ge t_0$;
\item $z(t)\in\Z$ and $u(t)\in\U$ $\ \forall t\ge t_0$;
\item $\lVert\yg(t)\rVert_\infty\le\Mg$ $\ \forall t\ge t_0$.
\end{enumerate}
\end{definition}

We now formally state the problem of interest:
\begin{problem}\phantomsection\label[problem]{prob:cert}
Given the closed-loop system \eqref{eq:loop}, the constraint requirement \eqref{eq:req}, and the I-O CBF Governor \eqref{eq:qpproj}, find a nonempty $\Sset\subseteq\Omega$ such that the I-O CBF Governor satisfies SIOCF on $\Sset$ for a given $\alpha^{1}$.
\end{problem}

%===========================================%
%   DEFINITIONS, PROPERTIES, AND THEOREMS   %
%===========================================%

\subsection{Definitions, Properties, and Theorems}\label{sec:solutions}
\subsubsection{Definitions}
We summarize a few key definitions below and refer the reader to \PartI\ for further details. First, we define the polyhedral cone
\begin{equation*}
    \Lambda:=\{\lambda\in\mathbb R^{6m}:A_\zeta^\top\lambda=0,\ \lambda\ge0\},
\end{equation*}
and let \(\{\lambda^1,\ldots,\lambda^N\}\) denote its normalized extreme rays. For any nonempty set \(\mathcal S\subseteq\mathbb R^{n_p+m}\), define
\[
\nu_k(\mathcal S):=\inf_{\bar x\in\mathcal S}\lambda^{k\top}b(\bar x),
\qquad k=1,\ldots,N.
\]
 Define
\[
Q:=
\begin{bmatrix}
C_z\\
C_z\bar A\\
\Kfull
\end{bmatrix}, \qquad
\bar C:=
\begin{bmatrix}
C_z\\
\Kfull
\end{bmatrix}.
\]
Define $\Oh$, which corresponds to the unobservable subspace of \((\bar A,\bar C)\), given by
\[
\mathcal O
:=
\bigcap_{k\ge0}
\ker(\bar C\bar A^k).
\]
Also choose $V\in\R^{n_p\times d}$ such that
\begin{equation}\label{eq:zeroconddef}
\operatorname{span}V=\ker\begin{bmatrix}\Clim\\ \Clim A_p\end{bmatrix},
\end{equation}
where $\operatorname{rank}V=d$.
We next introduce a nonsingular coordinate transformation $\xi=\Tz\xb$, where 
\begin{gather}\label{eq:coord1}
    \xi=[\,z\T\ \ \dot z\T\ \ \eta\T\,]\T,    
\end{gather}
and \(\eta\in\mathbb R^{n_p-m}\) denotes an internal state. From Lemma 1 in \PartI, it can be shown that $\eta$ is the state of a dynamic system given by
\begin{equation}
    \dot\eta=A_\eta\eta+B_d\dot z, \label{eq:etadyn}
\end{equation}
where \(A_\eta\) is Hurwitz. Also let $W_p\in\R^{(n_p-m)\times n_p}$ have rows spanning $\ker(B_p\T)$, and define 
\begin{equation}
    \etat:=W_px_p\in\R^{n_p-m}. \label{eq:etatilde}
\end{equation}

We define a compact convex set \(E\subset\mathbb R^{n_p-m}\), with \(0\in E\), to be {\em robustly positively invariant }(RPI) if
\[
\eta(t_0)\in E
\quad\Longrightarrow\quad
\eta(t)\in E,\qquad \forall t\ge t_0,
\]
for every admissible \(\dot z(t)\) satisfying the bounds imposed by \(\Omega\).  Such a set $E$ is explicitly defined as
\begin{equation}
    \begin{gathered}
    E:=\operatorname{cl}\Big\{\operatorname{conv}\Big\{e^{A_\eta(t-t_0)}\eta(t_0)
    +\int_{t_0}^{t}e^{A_\eta(t-\tau)}\Bd\,w(\tau)\,d\tau:\\
    t\ge t_0,\ w(\tau)\in\W\Big\}\Big\},\label{eq:add2}
    \end{gathered}
\end{equation}
with $\eta(t_0)\in\Pieta$, where $\W:=\{w\in\R^m:\ -\alpha_1w_z\le w\le\alpha_1w_z\}$ and $w_z:=\zmax-\zmin$, $\operatorname{conv}(\cdot)$ denotes the convex hull and $\Pieta:=\{\eta(\xb):\ \xb\in\Omega\}$. 

Finally, we introduce a design parameter vector $\alpha^2$ in order to accommodate a relaxation in the input constraint as 
\begin{equation}\label{eq:alpha2}
\alpha^{2}:=[\,\alpha_1\ \ \alpha_2\ \ \alpha_u\ \ \alpha_g\ \ \varrho\,]^\top .
\end{equation}
leading to the input constraint
\begin{equation}\tag{2a$'$}\label{eq:requrho}
u(t)\in\varrho\,\U:=\{u\in\R^m:\ \varrho\,\umin\le u\le\varrho\,\umax\}\ \ \forall t\ge t_0.
\end{equation}

%===========================================%
%                PROPERTIES                 %
%===========================================%

\subsubsection{Properties of the I-O CBF Governor}

The first set of properties we list below corresponds to Propositions 1, 3, and 4 in \PartI:

\begin{property}\label{prop:cert}
$F(\xb)\neq\varnothing\ \forall\xb\in\Sset$ if and only if
\begin{equation}\label{eq:siocfcond}
\min_k\nu_k(\Sset)\ge0 .
\end{equation}
\end{property}

\begin{property}\label{thm:fin}
$\forall\alpha^{1}\in\R_{++}^{4}, \
\nu_k\big(\Omega\big)>-\infty\ \ \forall k=1,\dots,N$ if and only if
\begin{equation}\label{eq:finitecond}
    \ker Q=\Oh.
\end{equation}
\end{property}

The inequality \eqref{eq:siocfcond} is denoted as the \textit{SIOCF condition} on \(\mathcal S\), and \eqref{eq:finitecond} as the \textit{finiteness condition}. 

\begin{property}\label{thm:zeroconddefV} 
If $\Ab$ satisfies \textnormal{(A3)}, $\ker Q=\Oh$ if and only if
\begin{equation}\label{eq:zerocond}
\begin{bmatrix}\Clim A_p^{2}\\ K_PA_p+K_I\Creg\end{bmatrix}V=0,
\end{equation}
where $V$ is given by \eqref{eq:zeroconddef}. 
\end{property}

With the above properties, we restate the key results from \PartI\ below.
\begin{theorem}\label{thm:mainSE}
Let $E$ be any RPI set of \eqref{eq:etadyn} with $\eta(t_0)\in E$, and define the compact set
\begin{equation}\label{eq:SEset}
\SE:=\Omega\cap \{\xb:\ \eta(\xb)\in E\} .
\end{equation}
Then the following hold:
\begin{enumerate}[label=(\alph*),leftmargin=2.2em]
\item Let the finiteness condition hold. Then, the I-O CBF Governor satisfies SIOCF on $\Omega$ if and only if 
the SIOCF condition is satisfied on $\Omega$;
\item the I-O CBF Governor satisfies SIOCF on $\SE$ if and only if
the SIOCF condition is satisfied on $S_E$.
\end{enumerate}
\end{theorem}

Since \Cref{thm:mainSE} requires a necessary and sufficient condition, a solution to the \Cref{prob:cert} need not exist for any choice of \(\alpha^1\). The following theorem from \PartI\ resolves this difficulty by expanding the design space of the I-O CBF Governor from $\alpha^1$ to $\alpha^2$. 

\begin{theorem}\label{thm:fix4}
$\forall(\alpha_1,\alpha_2,\alpha_u)\in\R_{++}^{3}$, $\exists\,\ag^{\star},\varrho^{\star}$ such that\\
the I-O CBF Governor satisfies SIOCF on $\SE(\varrho)$ $\forall(\ag,\varrho)\in(\ag^{\star},\infty)\times[\varrho^{\star},\infty)$.
\end{theorem}

%===========================================%
%        Implementation and Design          %
%===========================================%

\subsection{Implementation and Design}\label{sec:design}
The core of the I-O CBF governor constitutes the determination of the modified command $y_g$, which is obtained by solving the QP problem \eqref{eq:qpproj}. 
The online computational burden is comparable to that of a QP solver \cite{osqp}, whereas for SISO systems an exact closed-form solution is available. The offline computation consists primarily of checking the SIOCF condition, which scales as \(O(N)\) in the number of extreme rays \(N\), as elaborated below.

\Cref{thm:mainSE}(a) requires that we first check if the finiteness condition is satisfied, which can be accomplished by ensuring that \eqref{eq:zerocond} holds. Once the finiteness condition is satisfied, for the case of \Cref{thm:mainSE}(a), we  need to ensure that the SIOCF condition is satisfied on $\Omega$. For this evaluation, 
the normalized extreme rays \(\{\lambda^1,\ldots,\lambda^N\}\) of \(\Lambda\) are first computed. These rays are utilized to compute $\min_k\nu_k (\Omega)$,  where
\begin{equation}
    \nu_k(\Omega)
=\inf_{\bar x\in\Omega}\lambda^{k\top}b(\bar x)
\end{equation}
It should be noted that computation of $\nu_k$ for a given $\lambda^k$ is a linear program. In summary, checking the finiteness condition, followed by checking the SIOCF condition on $\Omega$ suffices to guarantee the existence of a solution for the I-O CBF Governor from \Cref{thm:mainSE}(a).

When the finiteness condition is not satisfied, one needs to determine an RPI set $E$ which satisfies \eqref{eq:etadyn}. For this purpose, we use \eqref{eq:add2} and a polyhedral approximation
of $E$ using the methods in \cite[Ch.~6]{BlanchiniMiani15}, whose cost is set by the desired accuracy, i.e., by the number of supporting hyperplanes, and is independent of $N$. This in turn allows the determination of $S_E$ as in \eqref{eq:SEset}. One can then check the SIOCF condition on $S_E$, which requires $N$ linear programs, one per extreme ray; by \Cref{thm:mainSE}(b), this guarantees a solution for the I-O CBF Governor.

We now proceed to the design of the I-O CBF Governor as outlined in \Cref{thm:fix4} and \cref{alg:fix} which includes the choice of the parameter vector $\alpha^2$. As in \Cref{thm:mainSE}, the online computation of the I-O CBF Governor is simply enabled by a QP solver. We discuss the offline implementation aspects below. 

We repeat Algorithm 1 here for the sake of completeness, but refer the reader to \PartI\ for details of all variables therein.
The procedure in \cref{alg:fix} can be summarized as follows: First, $\varrho$ is initialized with \(\varrho_1=1\), retaining the  input constraints as in \eqref{eq:req}, and
\begin{equation}\label{eq:a1min}
\alpha_1^{\min}:=\inf\big\{\alpha_1>0:\ \psi_1^{\pm}(\xb)\ge0\ \ \forall\xb\in X_0\big\},
\end{equation}
where \(X_0\) is the prescribed set of initial conditions. This choice ensures \(X_0\subseteq\Omega\).  
The algorithm then proceeds to select \(\alpha_1\), which determines the geometry of \(\Omega\) and \(S_E\), and finally selects \((\alpha_2,\alpha_u,\alpha_g)\) in Steps 1 and 2 to satisfy the SIOCF condition $\min_k\nu_k(S_E)\geq0$. If this is not possible, the procedure increases \(\varrho\) using Step 3 in \cref{alg:fix}.

\begin{algorithm}[t]
\caption{$\alpha^{2}$ design procedure, $\varrho_1:=1$}\label{alg:fix}
\noindent\hspace*{-1.7em}\begin{minipage}{\dimexpr\linewidth+1.7em\relax}
\begin{algorithmic}[1]\makeatletter\def\ALC@lno{}\makeatother
\STATE Compute the extreme rays of $\Lambda$ and $\alpha_1^{\min}$ as in \eqref{eq:a1min}.
\STATE Construct an RPI set $E$ and $\SE(\varrho_1)$.
\STATE \emph{(Step 1)} Compute $\ThH(\alpha_1^{\min},\varrho_1)$ as in \cite[Eq.~(29)]{YoonPartIACC}; if $\ThH\ge0$, compute $\ThB(\alpha_1,\varrho_1)$ as in \cite[Eq.~(30)]{YoonPartIACC} over $[\alpha_1^{\min},\bar\alpha_1]$.
%\STATE \emph{(Step 2)} Increase $\alpha_1$, and recompute $E$, $\SE(\varrho_1)$ and $\ThB(\alpha_1,\varrho_1)$.
\IF{$\ThH\ge0$ and $\ThB(\alpha_1,\varrho_1)>0$ for some $\alpha_1\in[\alpha_1^{\min},\bar\alpha_1]$}
\STATE \emph{(Step 2)} Set $\alpha_1=\alpha_1^{\star}$ as defined in \cite[Eq.~(34)]{YoonPartIACC}, choose $\varepsilon\in(0,\ThB)\cap\R$ and $c>0$.
\STATE Set $\alpha_2\ge\underline\alpha_2$ with $\alpha_u=c\,\alpha_2$, and $\ag\ge\underline\alpha_g(\alpha_2)$ defined in \cite[Eq.~(37)]{YoonPartIACC} and \cite[Eq.~(38)]{YoonPartIACC}.
\ELSE
\STATE \emph{(Step 3)} Choose $\varepsilon\in[0,\nub]$, and set $\ag>\ag^{\star}$ and $\varrho\ge\varrho^{\star}$ defined in \cite[Eq.~(23)]{YoonPartIACC}.
\ENDIF
\end{algorithmic}
\end{minipage}
\end{algorithm}

%===========================================%
%            NUMERICAL EXAMPLES             %
%===========================================%

\section{Numerical Examples}\label{sec:example}
\begingroup
\renewcommand{\thesubsectiondis}{Example~\arabic{subsection}:}
\renewcommand{\thesubsection}{\arabic{subsection}}
\crefalias{subsection}{example}

In this section, we illustrate the main theoretical results established in \PartI\ through six numerical examples. Examples 1 and 2 demonstrate \Cref{thm:mainSE}(a) with the I-O CBF Governor through %command tracking and constraint satisfaction 
simulations of high performance flight platforms for a command tracking task. {Example 1 corresponds to flight control of the F-4 MIMO lateral--directional dynamics, while Example 2 corresponds to the X-29A longitudinal short--period dynamics, which are open--loop unstable.} In both examples, we demonstrate reasonable tracking and I-O constraint satisfaction. %However, \Cref{thm:mainSE}(a) no longer applies and when the finiteness condition is violated. Instead, \Cref{thm:mainSE}(b) provides a procedure for constructing a solution to Problem 2, shown in 
Example 3 is an academic one where the finiteness condition is not satisfied. It is shown first that $\Omega$ is no longer forward invariant. We then proceed to construct a  set $E$ leading to a smaller forward invariant set $S_E$, where once again reasonable tracking and I-O constraint satisfaction are demonstrated, thereby corroborating \Cref{thm:mainSE}(b). %Under infeasible constraints that render \Cref{thm:mainSE} inapplicable, \Cref{thm:fix4} permits input constraint relaxation while respecting output constraints as illustrated in Example 4. 
Example 4 sets the stage for \Cref{thm:fix4}, where we first begin with an example where for a specific choice of $\alpha^1$, the SIOCF condition is not satisfied for a set $S_E$. We then proceed to show that for an expanded design parameter $\alpha^2$, the SIOCF condition can be satisfied. Example 5 provides a numerical demonstration of \cref{alg:fix}, %used for the I-O CBF Governor design, 
which leads to a more efficient choice of $\alpha^2$. Finally, Example 6 highlights the importance of condition (A2) in \Cref{ass:standing}. All examples consist of plant dynamics whose order is greater than or equal to two to illustrate the properties of the I-O CBF Governor. 
A closed-form solution exists for the I-O CBF Governor for all Examples 2 to 6, while Example 1 requires a QP solver.
We also refer the reader to an interesting example of an open-loop unstable scalar plant in \PartI\ where it is shown that no controller can be found that allows the I-O constraints to be satisfied.

We note that the QP in \eqref{eq:qpproj} may become infeasible in some cases. The procedure we have adopted here is to relax the QP constraints in such cases, so that the simulation continues and does not terminate. This relaxation consists of a procedure where the bound on $\yg$ is relaxed first, and if necessary, the output CBF constraints are relaxed as well until QP has a solution.

%===========================================%
%                Example 1                  %
%===========================================%

\subsection{Illustration of \Cref{thm:mainSE}(a)}\label{sec:exa}

Consider a lateral-directional model of the McDonnell F-4 Phantom fighter aircraft (see \cite{LavretskyWise}, Chapter 14.8), with plant states $x_p=\begin{bmatrix}\beta & p & r\end{bmatrix}^{\top}\in\R^{3}$ and control inputs $u=\begin{bmatrix}\delta_a & \delta_r\end{bmatrix}^{\top}\in\R^{2}$, where $\beta$ is the sideslip angle (rad), $p$ and $r$ are the roll and yaw rates (rad/sec), and $\delta_a$ and $\delta_r$ are the aileron and rudder deflections (rad). The plant dynamics in \eqref{eq:plant} are defined by
\begin{gather}
A_p=\begin{bmatrix}-0.11794&0.00085&-1.00010\\ -7.01130&-1.44920&0.22059\\ 6.30350&0.06511&-0.41172\end{bmatrix}, \\
B_p=\begin{bmatrix}0&0.015257\\ -7.96620&2.68750\\ 0.60926&-2.35770\end{bmatrix}.
\end{gather}
The regulated output $y=\begin{bmatrix}p & n_y\end{bmatrix}^{\top}$ consists of the roll rate $p$ and lateral-load-factor $n_y$, expressed in units of $g$. The output matrices in \eqref{eq:plant} are
\begin{gather}
\Creg=\begin{bmatrix}0&1&0\\ -2.6049&0.018724&0.067695\end{bmatrix},\\
\Dreg=\begin{bmatrix}0&0\\ 0&0.33698\end{bmatrix}.
\end{gather}
The output in \eqref{eq:zdef} is chosen as $z=\begin{bmatrix}\beta & p\end{bmatrix}^{\top}$ with
\begin{gather}
    \Clim= \begin{bmatrix} 1 & 0 & 0\\ 0 & 1 & 0 \end{bmatrix}.
\end{gather}
The controller gains
\begin{gather}
    K_P=\begin{bmatrix}0.8699&-0.9686&-0.4135\\ 1.4680&0.3352&-1.3832\end{bmatrix},\\
    K_I=\begin{bmatrix}-1.2220&-0.3298\\ 0.4705&-1.7434\end{bmatrix},
\end{gather} 
are obtained from an LQR-based design with
\begin{gather}
    Q=\text{diag}\left(\begin{bmatrix}1.025&1.0289&0&1.6021&1.6021\end{bmatrix}\right),\\
    R=\text{diag}\left(\begin{bmatrix}1&0.49129\end{bmatrix}\right).
\end{gather}
The I-O constraints are specified as in~\cite{Menner_2026}
\begin{gather}
    \U=[-3^{\circ},3^{\circ}]\times[-2^{\circ},2^{\circ}], \\
    \Z=[-0.5^{\circ},0.5^{\circ}]\times[-18^{\circ}/\mathrm{s},18^{\circ}/\mathrm{s}].
\end{gather}

The free parameters of the I-O CBF Governor in this simulation were determined by Steps 1-2 in \cref{alg:fix} as $\alpha^{1}=(0.7,21.2,212.2,5.70)$. Smooth commands $\ycmd$ were chosen, with  \Cref{fig:exa_traj}(a) illustrating the tracking outputs $y$, \Cref{fig:exa_traj}(b) the outputs $z$, and the inputs $u$ in \Cref{fig:exa_traj}(c). It is clear from these figures that tracking performance is satisfactory and I-O constraint satisfaction is met as well. Although the spikes in \(y_g\) can be reduced through a more appropriate choice of \(\alpha^1\), no such tuning is performed here.

\begin{figure}
    \centering
    \begin{subfigure}{1\columnwidth}
        \centering
        \includegraphics[width=\linewidth]{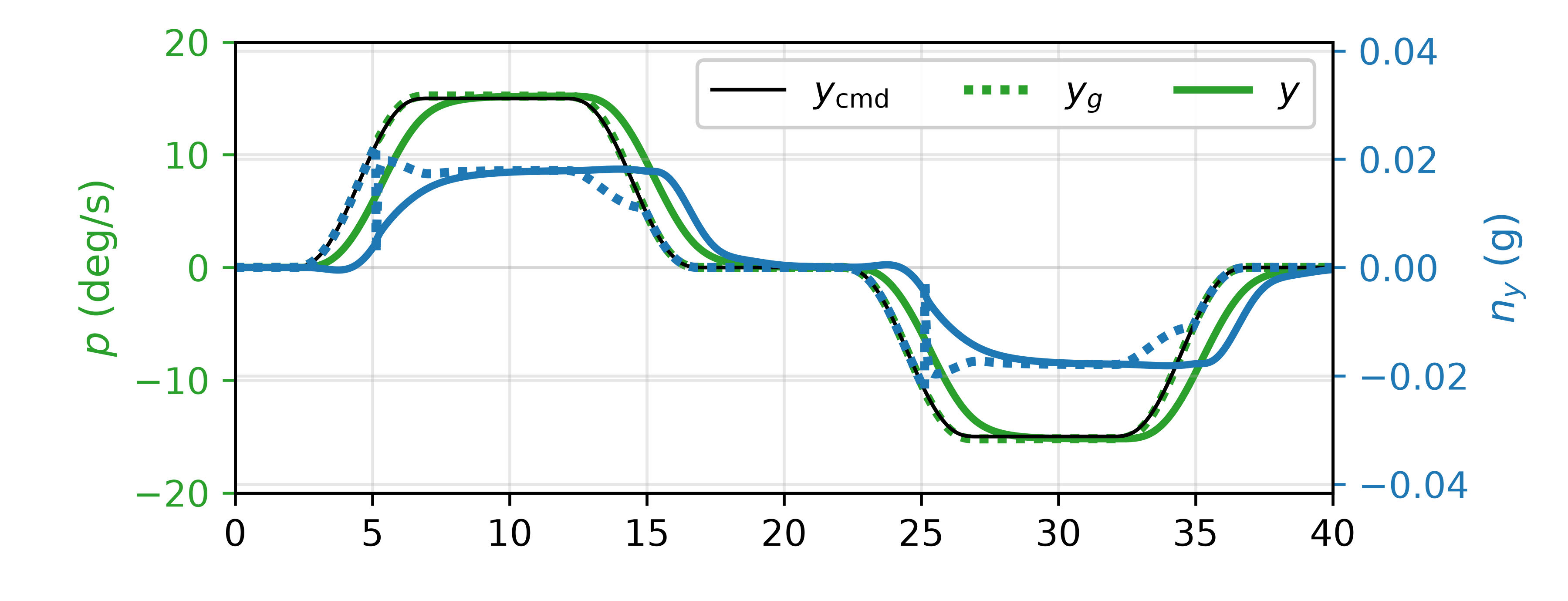}
        \vspace{-7mm}
        \caption{Command tracking}
        \label{fig:exampleA_tracking}
    \end{subfigure}
    \begin{subfigure}{1\columnwidth}
        \centering
        \includegraphics[width=\linewidth]{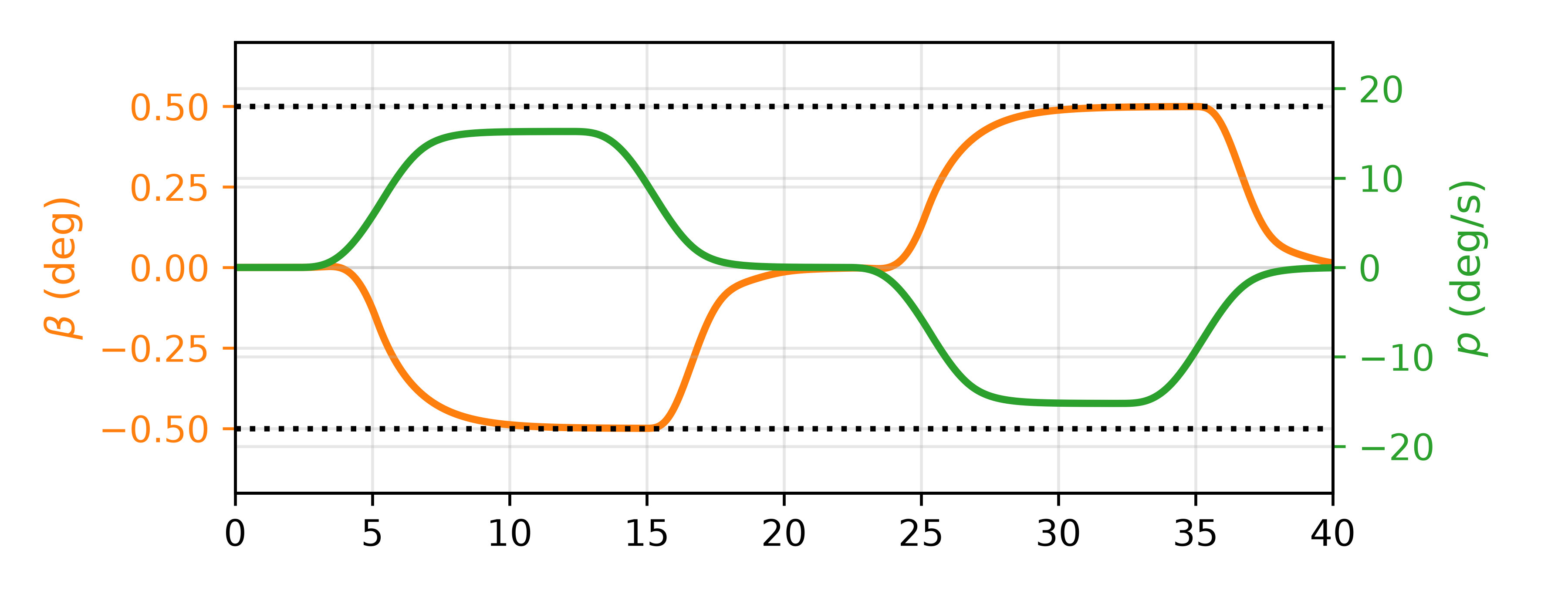}
        \vspace{-7mm}
        \caption{Outputs $z$}
        \label{fig:exampleA_output}
    \end{subfigure}
    \begin{subfigure}{1\columnwidth}
        \centering
        \includegraphics[width=\linewidth]{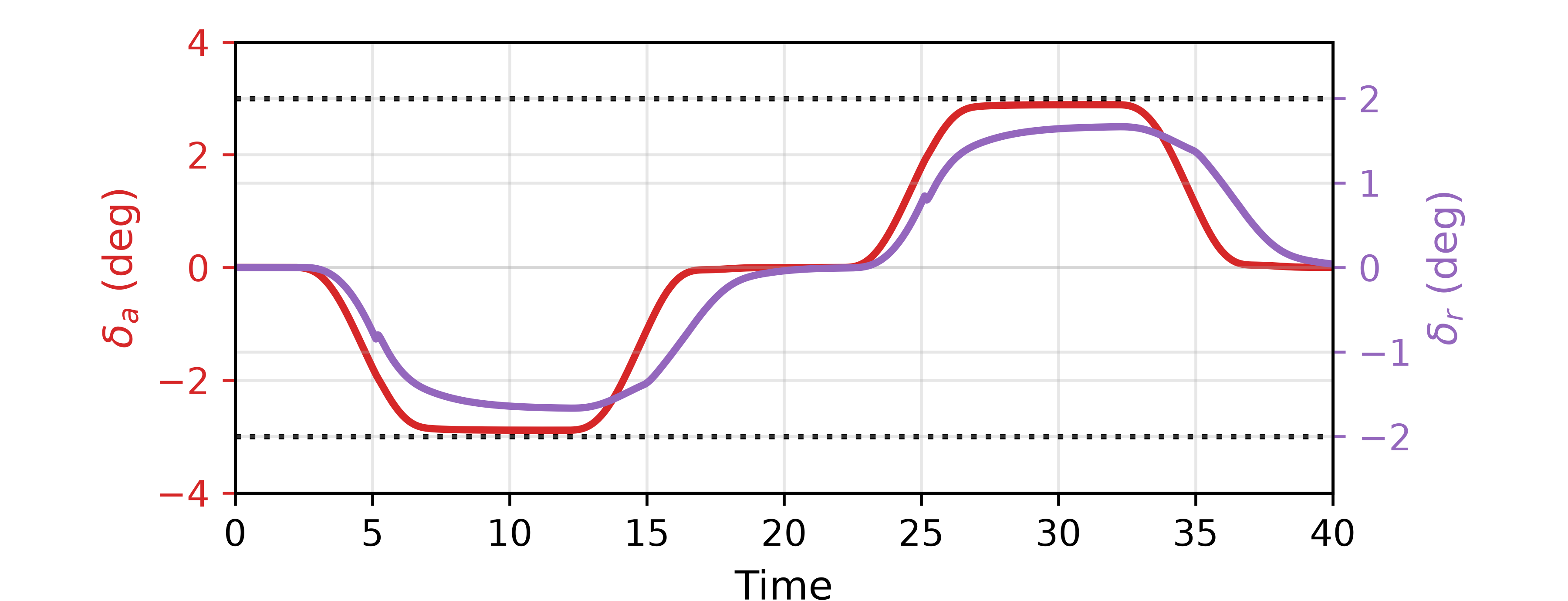}
        \vspace{-4.5mm}
        \caption{Inputs $u$}
        \label{fig:exampleA_input}
    \end{subfigure}
    \caption{Closed-loop system performance of  McDonnell F-4 Phantom fighter aircraft lateral-directional model under the I-O CBF Governor.}
    \label{fig:exa_traj}
\end{figure}

A further examination of these figures shows the effect of the I-O constraints on command tracking. For the roll rate, the governed reference remains close to the command, resulting in close tracking of $p$. In contrast, the governed lateral-load-factor reference is significantly reduced relative to its command, resulting in the observed tracking error in $n_y$. This modification reflects a fundamental conflict between command following and constraint satisfaction. The governor resolves this conflict by modifying the reference rather than permitting a constraint violation. At \(p=15^\circ/\mathrm{s}\), a steady-state constrained-feasibility calculation gives a maximum admissible lateral-load-factor of \(0.0178g\), compared with the commanded \(0.0312g\). Thus, at steady state, increasing \(n_y\) further while maintaining the commanded roll rate \(p\) would require violation of at least one prescribed input or output limit.

Additional dissection of the I-O CBF Governor is carried out below. First, satisfaction of (A1)-(A3) in \Cref{ass:standing} is numerically verified: $\det(H_u)=0.1215$, the invariant zero of  $(A_p,B_p,\Clim,0_{m\times m})$ is at $ -141.5\in\mathbb{C}^{-}$, and the eigenvalues of $\Ab$ are $\sigma(\Ab)=\left\{ -8.83, -1.05\pm2.45j, -1.24, -0.85 \right\}$. Next, the finiteness condition is verified using \Cref{thm:zeroconddefV} by noting that $d=0$. Finally, the SIOCF condition was observed to hold since $\min_k\nu_k(\Omega)=0.16$.

%===========================================%
%                Example 2                  %
%===========================================%

\subsection{Illustration of \Cref{thm:mainSE}(a) with SISO System}\label{sec:exf}

Consider a longitudinal model of the Grumman X-29A aircraft (see \cite{bosworth1992linearized} and \cite{LavretskyWise}, Chapter 14.7),
with plant states $x_p=\begin{bmatrix}\alpha & q\end{bmatrix}^{\top}\in\R^{2}$ and control input $u=\delta_e\in\R$, where $\alpha$ is the angle of attack (rad), $q$ is the pitch rate (rad/sec), and $\delta_e$ is the elevator deflection (rad). The plant dynamics in \eqref{eq:plant} are defined by
\begin{gather}
A_p=\begin{bmatrix}-2.24&0.9897\\ 44.74&-0.902\end{bmatrix},\qquad B_p=\begin{bmatrix} -0.233\\ -45.93\end{bmatrix}
\end{gather}
The regulated output matrices for $C^*$ command tracking are
\begin{gather}
\Creg=\begin{bmatrix}66.5&12.745\end{bmatrix}, \qquad \Dreg=8.5863.
\end{gather}
The output in \eqref{eq:zdef} is chosen as
\begin{gather}
    \Clim= \begin{bmatrix} 0&1 \end{bmatrix},
\end{gather}
and the controller gains are
\begin{gather}
    K_P=\begin{bmatrix}-1.4472&-0.24008\end{bmatrix},\qquad K_I=-0.01.
\end{gather} 
The I-O constraint sets in \eqref{eq:req} are 
\begin{gather}
    \U=[-6^{\circ},6^{\circ}], \qquad \Z=[-16^{\circ}/\mathrm{s},16^{\circ}/\mathrm{s}].
\end{gather}
Although \((A_p,B_p,\Creg,\Dreg)\) has the RHP invariant zero at \(+74.42\) and the open-loop plant is unstable with $\sigma(A_p)=\{-8.259,\;5.117\}$, the system satisfies \Cref{ass:standing}.

For I-O CBF Governor parameters produced by Steps 1-2 in \cref{alg:fix}, $\alpha^{1}=(8.25, 2.464, 2.464, 20.912)$, \Cref{fig:ex6}(a) demonstrates satisfactory tracking of a smooth command while the outputs $z$ and the inputs $u$ reach and remain on their prescribed limits as shown in \Cref{fig:ex6}(b). It is straightforward to verify that this system satisfies the conditions in \Cref{ass:standing}. Furthermore, the finiteness condition is satisfied with $d=0$, and the necessary and sufficient condition $\min_k\nu_k(\Omega)=0.0075$.

For this unstable open-loop plant, an elevator magnitude of at least \(\delta_e=-\zmax/\Clim A_p^{-1}B_p=5.967^\circ\) is required to maintain the steady-state output boundary \(q=16^\circ/\mathrm{s}\). Thus, under tighter input constraints, the QP loses feasibility and constraints are subsequently violated.

\begin{figure}
    \centering
    \begin{subfigure}{1\columnwidth}
        \centering
        \includegraphics[width=\linewidth]{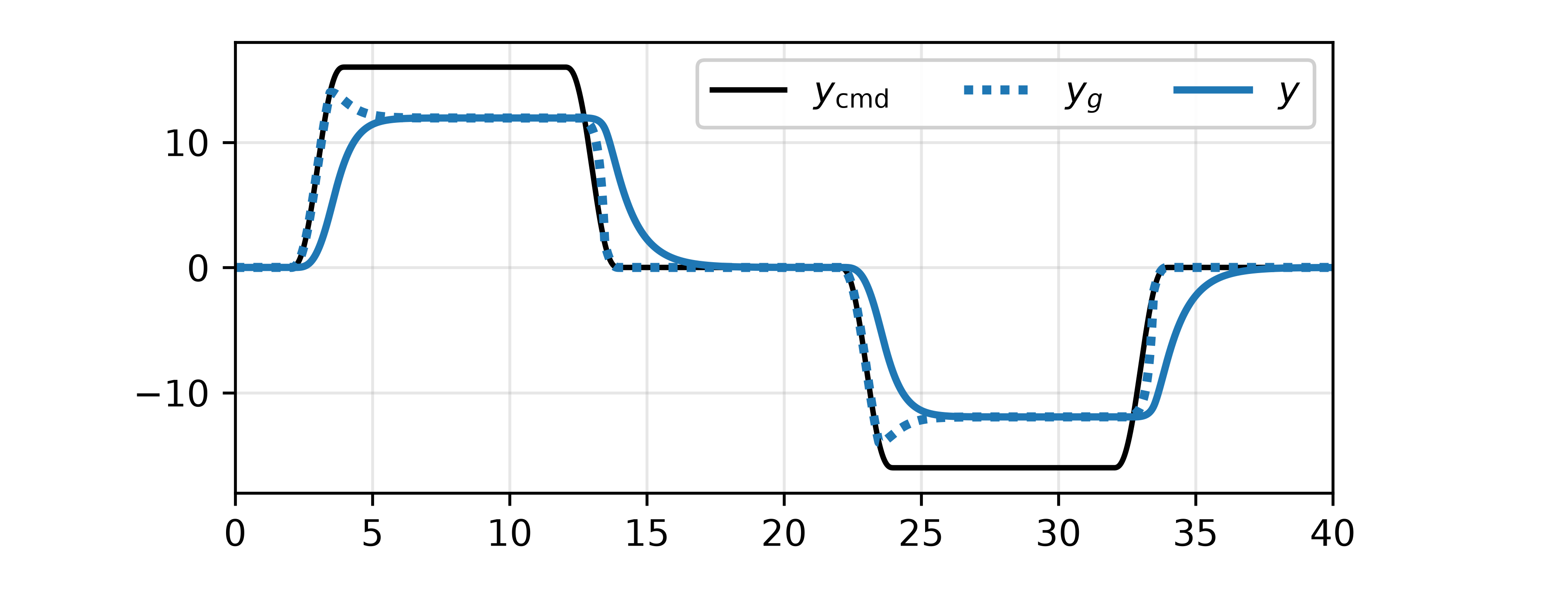}
        \vspace{-7mm}
        \caption{Command tracking}
    \end{subfigure}
    \begin{subfigure}{1\columnwidth}
        \centering
        \includegraphics[width=\linewidth]{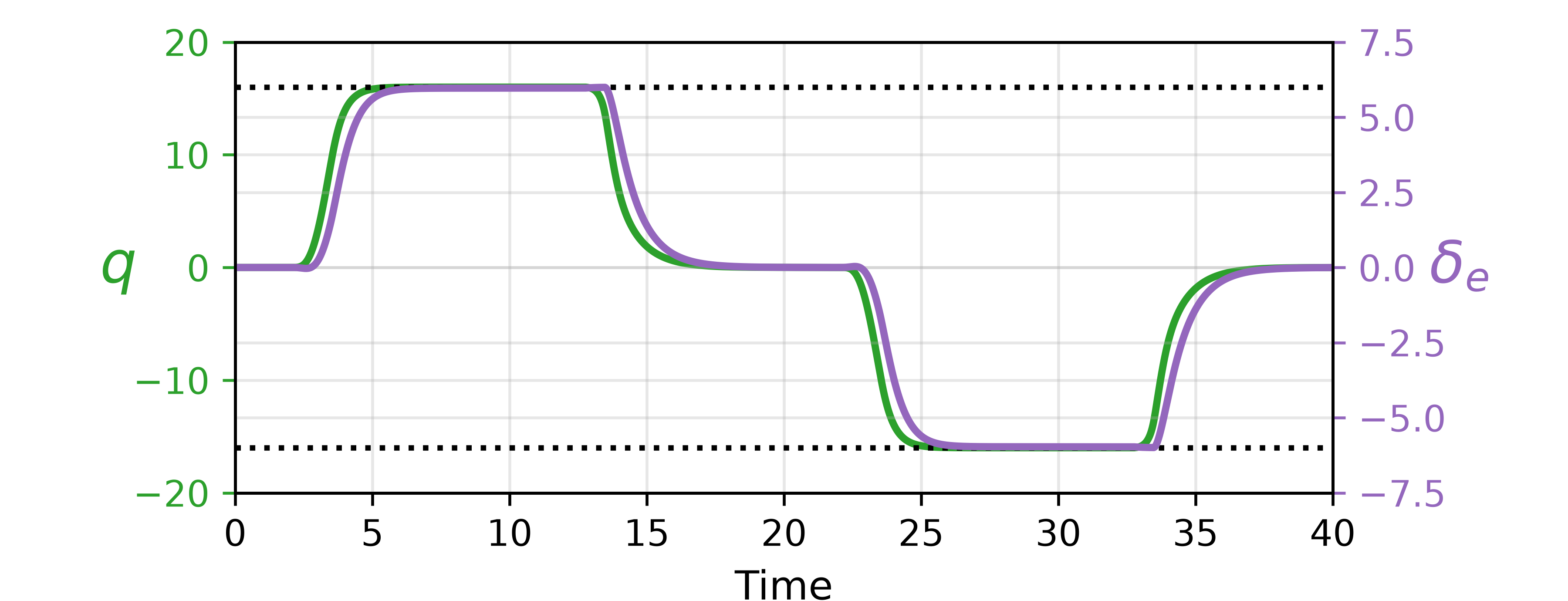}
        \vspace{-4.5mm}
        \caption{Output $z$ and input $u$}
    \end{subfigure}
    \caption{Closed-loop system performance of Grumman X-29A aircraft longitudinal model under the I-O CBF Governor.}
    \label{fig:ex6}
\end{figure}

%===========================================%
%                Example 3                  %
%===========================================%

\subsection{Illustration of \Cref{thm:mainSE}(b)}\label{sec:exb}

Consider a third-order SISO plant with state $x_p=\begin{bmatrix}x_1 & x_2 & x_3\end{bmatrix}^{\top}\in\R^{3}$ and input $u\in\R$. The plant dynamics in \eqref{eq:plant} are defined as
\begin{equation}
A_p=\begin{bmatrix}-1&2&0\\ 0&-2&3\\ 0&0&-3\end{bmatrix},\quad B_p=\begin{bmatrix}1\\0\\0\end{bmatrix}.
\end{equation}
The regulated output matrices in \eqref{eq:plant} are
\begin{equation}
    \Creg=
    \begin{bmatrix}
    1&0.5 &0
    \end{bmatrix},
    \qquad
    \Dreg=0.
\end{equation}
The output in \eqref{eq:zdef} is chosen as
\begin{equation}
    \Clim= \begin{bmatrix}
    1&0&0
    \end{bmatrix}.
\end{equation}
The controller gains are
\begin{equation}
    K_P=\begin{bmatrix}2&0&0\end{bmatrix},
    \qquad
    K_I=1,
\end{equation}
and the I-O constraint sets in \eqref{eq:req} are $\U=\Z=[-1,1]$.
It is straightforward to verify that the system satisfies \Cref{ass:standing}. 

It is easily seen that in this example, $\ker Q\supsetneq\mathcal \Oh$, and therefore that the finiteness condition cannot be satisfied for any $\alpha^{1}$. This in turn leads to the response in \Cref{fig:exb_traj}(a), where the I-O CBF Governor is implemented with $\alpha^{1}=(0.7, 19.1, 19.1, 9.91)$, produced by \cref{alg:fix}, and initialized at $\xb(0)=(0,0,5,0)\in\Omega$. For $\ycmd\equiv2$, the governor QP is infeasible at the initial time, i.e. no modified reference $\yg$ simultaneously satisfies all I-O CBF Governor QP constraints. Relaxing the QP constraints for visualization, the trajectory subsequently exits $\Omega$ and violates the output limits. This corroborates Proposition 2 in \PartI.

We demonstrate that \Cref{thm:mainSE}(b) holds by constructing $S_E$ as follows. We first choose the internal state $\eta$ from \eqref{eq:etadyn} as 
\begin{gather}
    \eta=\begin{bmatrix}x_2&x_3\end{bmatrix}^{\top}, \quad \dot{\eta}=A_\eta \eta, \quad A_\eta=\begin{bmatrix}-2&3\\ 0&-3\end{bmatrix}.
\end{gather}
with bounded initial conditions $\eta(0)=\begin{bmatrix} 0 & x_3(0)\end{bmatrix}^{\top}$ satisfying $|x_3(0)|\le\overline X$. The set \(E(\overline X)\) is constructed as the convex hull of the forward trajectories \(e^{A_\eta t}\eta_0\) generated from all such initial conditions for \(t\ge0\). Numerically, \(E(\overline X)\) is represented by a polyhedral outer approximation constructed from 360 uniformly distributed support directions. Finally, we obtain $S_E(\varrho,\overline X)=\Omega(\varrho)\cap\{\bar x:\eta(\xb)\in E(\overline X)\}$.

For initial condition $\xb(0)=(0,0,1.5,0)\in S_E(\varrho,\overline X)$ for $\overline X=1.5$ and $\varrho=1$, and under the same I-O CBF Governor parameters $\alpha^1$, \Cref{fig:exb_traj}(b) displays that the outputs and inputs respect their constraints.  This is further illustrated in \cref{fig:ss}, where some initial conditions in $E(5)$ may lead to constraint violations. In contrast, all initial conditions within $E(1.5)$ satisfy the I-O constraints. A numerical evaluation leads to $\min_k\nu_k(S_E)=1.00$, and therefore the I-O CBF Governor is SIOCF on $S_E$.

\begin{figure}
    \begin{subfigure}{1\columnwidth}
        \centering
        \includegraphics[width=\linewidth]{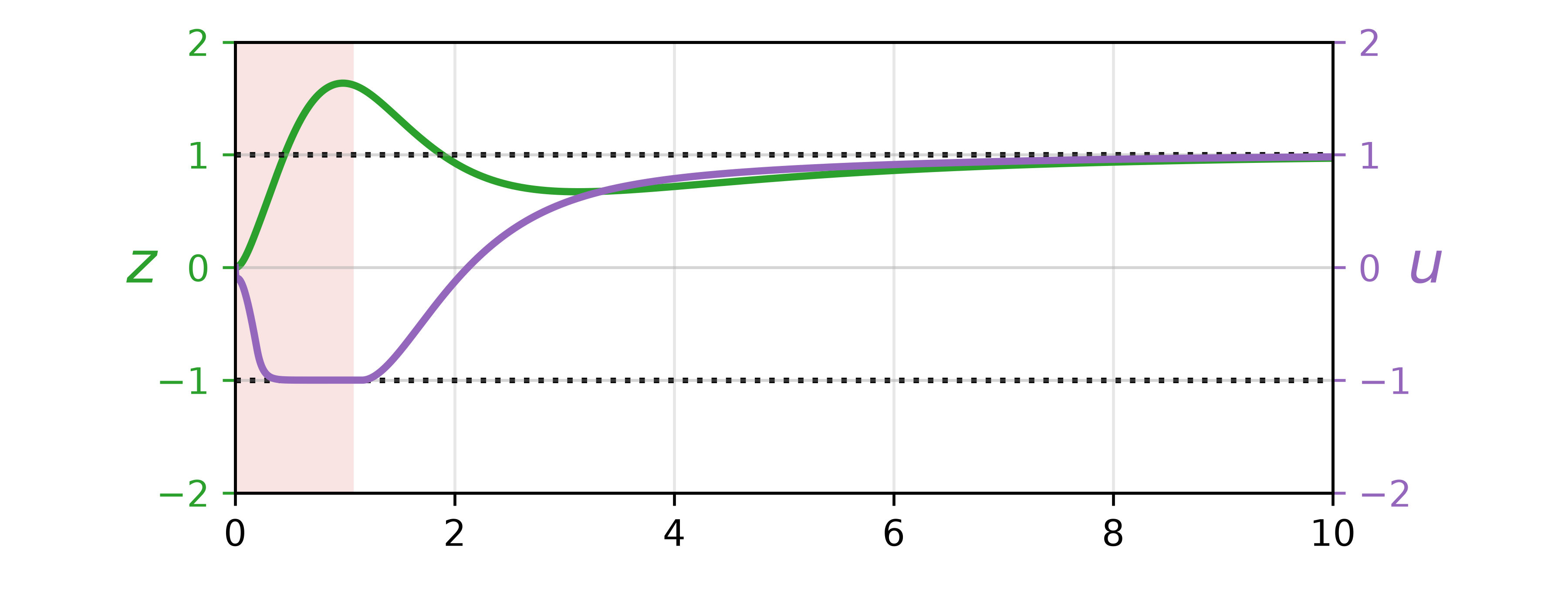}
        \vspace{-7mm}
        \caption{Constraints violated}
    \end{subfigure}
    \begin{subfigure}{1\columnwidth}
        \centering
        \includegraphics[width=\linewidth]{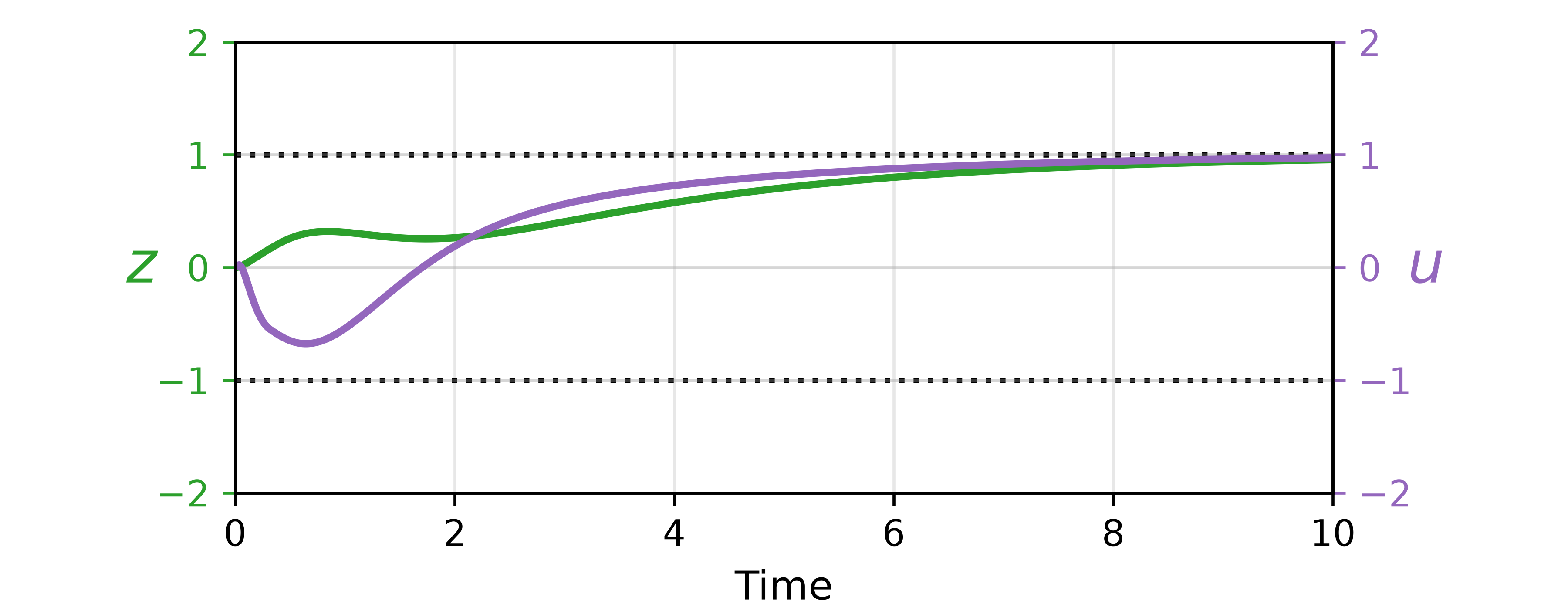}
        \vspace{-4.5mm}
        \caption{Constraints satisfied}
    \end{subfigure}
    \caption{Constraints are violated for $\overline X=5$ (top) and are satisfied for $\overline X=1.5$ (bottom).}
    \label{fig:exb_traj}
\end{figure}

\begin{figure}
    \centering
    \begin{subfigure}{1\columnwidth}
        \centering
        \includegraphics[width=\linewidth]{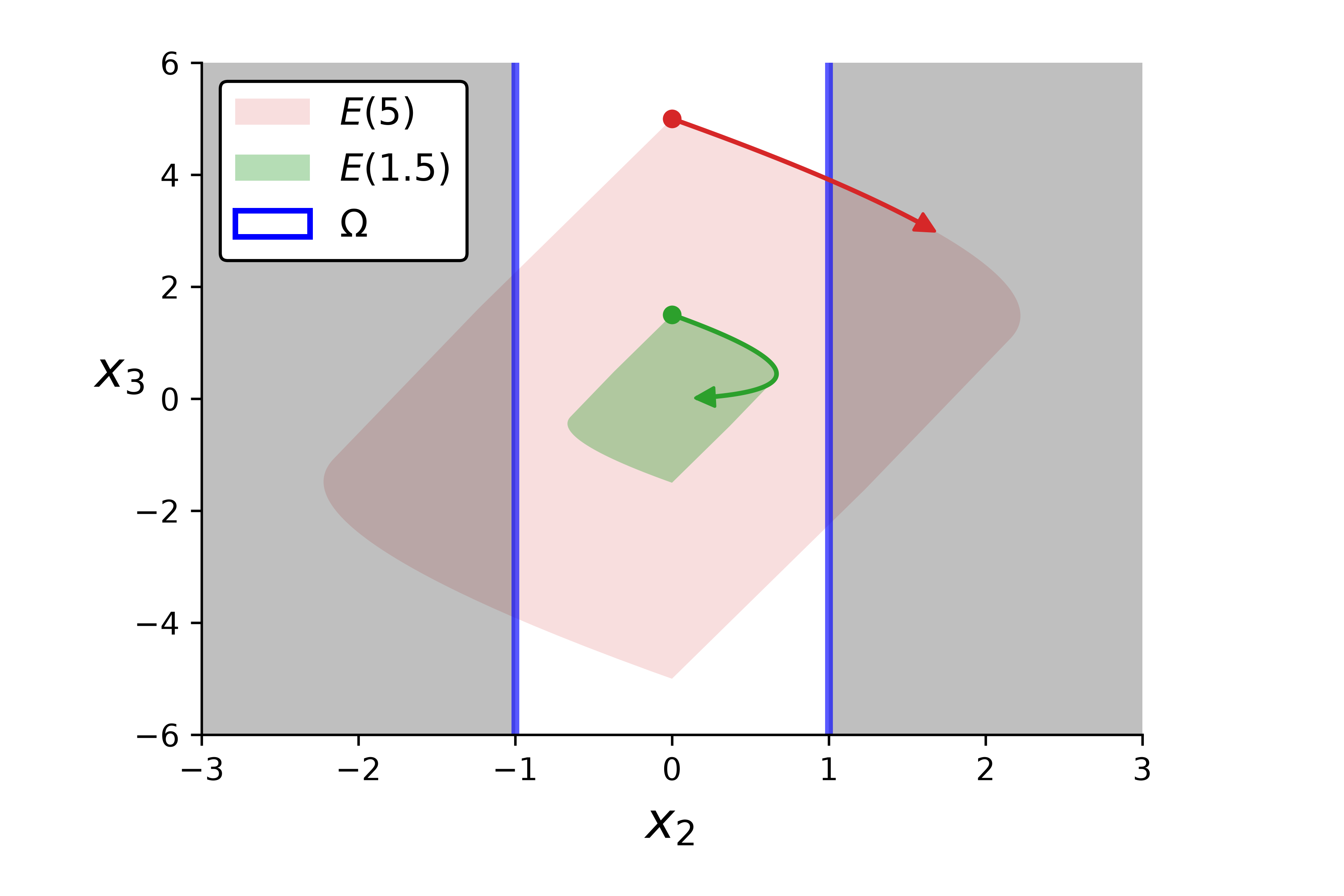}
        \vspace{-7mm}
    \end{subfigure}
    \caption{Initial internal states $\eta(0)$ in $E(5)$ may lead to constraint violations (red region), whereas $\eta(0)$ in $E(1.5)$ cannot (green region).}
    \label{fig:ss}
\end{figure}

%===========================================%
%                Example 4                  %
%===========================================%

\subsection{Illustration of \Cref{thm:fix4}}\label{sec:exc}

We consider the plant, controller gains, and constraints from \Cref{sec:exb}, but examine a larger set \(S_E\) constructed using $\overline X=5$. Importantly, there exist compact sets \(S_E\) for which no choice of $\alpha^1$ results in a SIOCF closed-loop system. Given the initial condition \(\bar x(0)=(0,0,5,0)\in S_E(1,5)\) and $\ycmd\equiv1$, the dashed lines in \Cref{fig:exc_rhostar} illustrate that $S_E(1,5)$ is not SIOCF for \(\alpha^1=(0.5,10,10,5)\). The I-O CBF Governor QP is infeasible at the initial time, and with the QP constraint relaxation, $z$ leaves $\Z$ at $t=0.437$. 

We now proceed to illustrate \Cref{thm:fix4} with the parameters $\alpha^2=(0.5,10,10,58.4, 5.84)$
produced by Step 3 of \cref{alg:fix}. For the same set of initial conditions as above, the solid lines in
\Cref{fig:exc_rhostar} show the corresponding output and relaxed input limits are respected under $\alpha^2$. 

In fact, the choice of $x_3(0)=5$ necessarily leads to a violation of the upper bound on $z$. For any input $|u(t)|\le\umax$ for all $t\geq0$, observe that
\begin{gather}
    z(t)\ge
   \overline X\,\Clim e^{A_pt}e_3
    -
    \umax\int_0^t
    |\Clim e^{A_ps}B_p|\,ds,
\end{gather}
where $e_3=[0,0,1]^{\top}$. The right hand side exceeds the upper bound on $z$ for any $\overline X \geq 4.5$. Another cross check for \Cref{thm:fix4} is the satisfaction of the SIOCF condition with $\alpha^2$ by computing $\min_k\nu_k(S_E(5.84,5))$ which gives 7.07.

\begin{figure}
    \centering
    \begin{subfigure}{1\columnwidth}
        \centering
        \includegraphics[width=\linewidth]{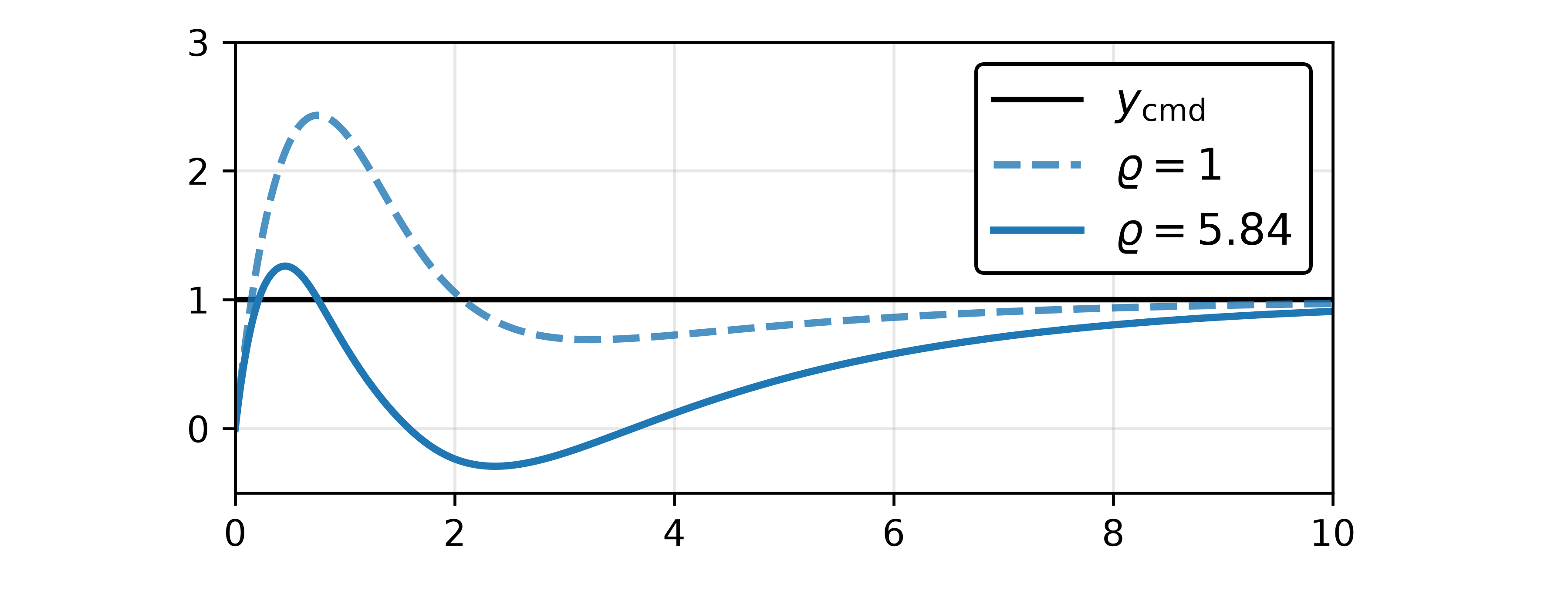}
        \vspace{-7mm}
        \caption{Command tracking}
    \end{subfigure}
    \begin{subfigure}{1\columnwidth}
        \centering
        \includegraphics[width=\linewidth]{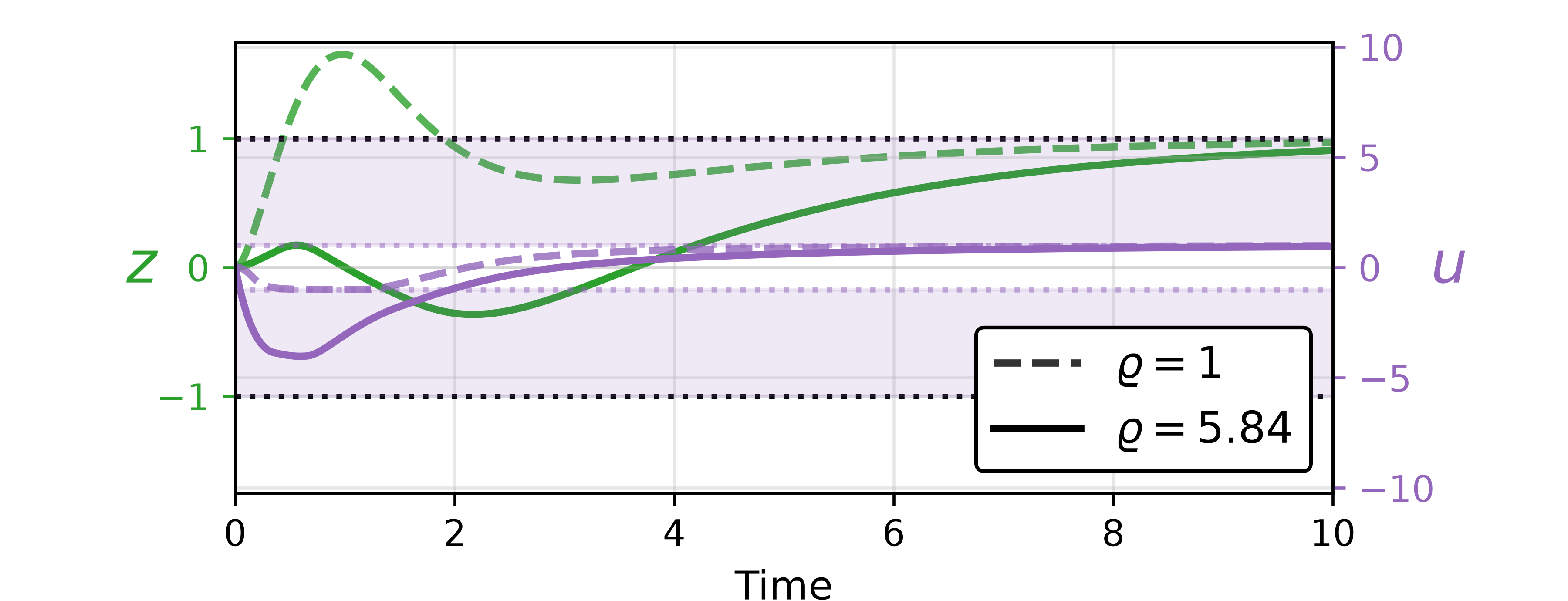}
        \vspace{-4.5mm}
        \caption{Output $z$ and input $u$}
    \end{subfigure}
    \caption{Increasing $\varrho$ enables satisfaction of output and relaxed input constraints.}
    \label{fig:exc_rhostar}
\end{figure}

%===========================================%
%                Example 5                  %
%===========================================%

\subsection{Illustration of \cref{alg:fix}}\label{sec:exd}

Given the plant, controller gains, and constraints from \Cref{sec:exb}, an efficient I-O CBF Governor design is obtained for $\overline X=1.5$ and initial parameters $(\alpha_1,\alpha_2,\alpha_u)=(0.5, 10, 10)$. For comparison, without tuning the given parameters, Step 3 of \cref{alg:fix} produces $\alpha^{2}=(0.5, 10, 10, 24.2,2.42)$, for which $\min_k\nu_k(S_E(2.42,1.5))=7.07$. The closed-loop behavior of the untuned parameters is illustrated by the dashed lines in \Cref{fig:exd_runs}. On the other hand, Steps 1-2 give $\alpha^{2}=(0.7,19.1,19.1,9.91,1)$ with $\varrho=1$, satisfying the SIOCF condition with \(\min_k\nu_k(S_E(1,1.5))=1.00\). Thus the input constraints in \eqref{eq:requ} are satisfied, as depicted by the solid lines in \Cref{fig:exd_runs}.

\begin{figure}
    \centering
    \begin{subfigure}{1\columnwidth}
        \centering
        \includegraphics[width=\linewidth]{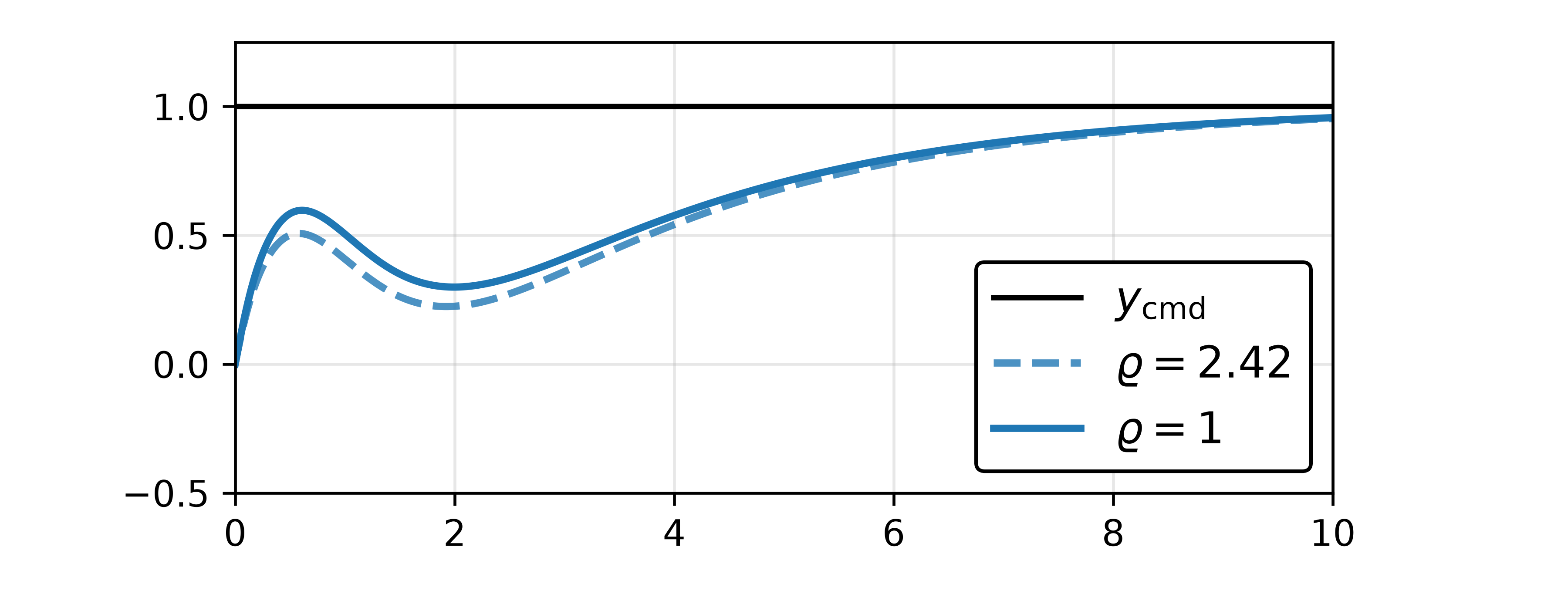}
        \vspace{-7mm}
        \caption{Command tracking}
    \end{subfigure}
    \begin{subfigure}{1\columnwidth}
        \centering
        \includegraphics[width=\linewidth]{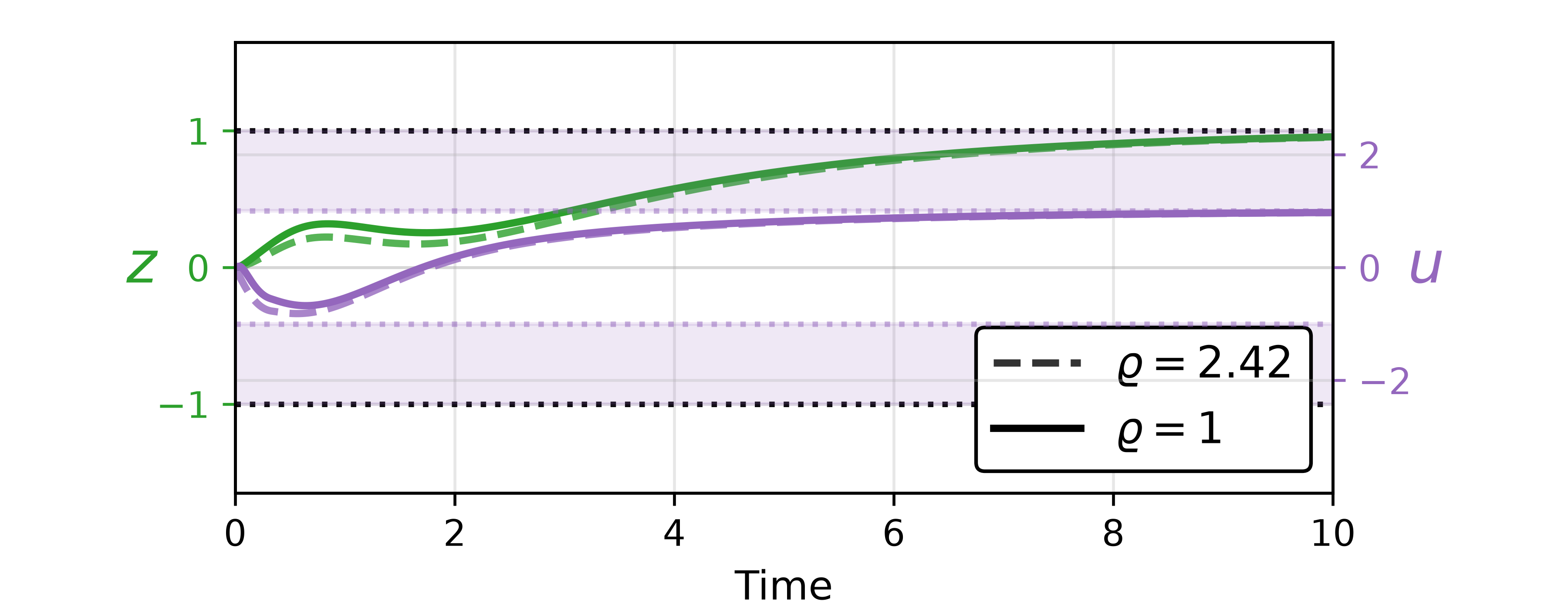}
        \vspace{-4.5mm}
        \caption{Output $z$ and input $u$}
    \end{subfigure}
    \caption{The design from Steps 1–2 of \cref{alg:fix} satisfies the constraints with lower control effort given $\xb(0)=(0,0,1.5,0)$.}
    \label{fig:exd_runs}
\end{figure}

%===========================================%
%                Example 6                  %
%===========================================%

\subsection{Violation of (A2) in \Cref{ass:standing}}\label{sec:exe}

Consider a second-order SISO plant with state $x_p\in\R^{2}$ and control input $u\in\R$. The plant dynamics in \eqref{eq:plant} are
\begin{equation}\label{eq:ex6plant}
A_p=\begin{bmatrix}0&1\\-2&-3\end{bmatrix},\qquad B_p=\begin{bmatrix}0\\1\end{bmatrix}.
\end{equation}
The output matrices in \eqref{eq:plant} and \eqref{eq:zdef} are
\begin{equation}\label{eq:ex6out}
\Creg=\begin{bmatrix}1 & 0.2\end{bmatrix}, \quad \Dreg=0, \quad \Clim= \begin{bmatrix} -1 & 1 \end{bmatrix}.
\end{equation}
The controller gains are
\begin{equation}\label{eq:ex6control}
    K_P=\begin{bmatrix}1.3 & 0.606\end{bmatrix},\qquad K_I=1,
\end{equation} 
and the input and output sets in \eqref{eq:req} are 
\begin{gather}
    \U=[-6,6], \qquad \Z=[-1,1].
\end{gather}
For these values, 
\((A_p,B_p,C_{\rm lim},0)\) has a RHP invariant zero of \(+1\).

It is easy to see that the closed loop system \eqref{eq:ex6plant}-\eqref{eq:ex6control} satisfies conditions (A1) and (A3) in \Cref{ass:standing}, but not (A2). A consequence of the RHP invariant zero implies that the inherent internal dynamics in this example is of the form
\begin{gather}\label{eq:ex6coord1}
    \dot{\tilde\eta}=\tilde\eta-z, \qquad
    \tilde\eta=-x_1,
\end{gather}
which can be derived using \eqref{eq:coord1}-\eqref{eq:etatilde} for the numerical values in \eqref{eq:ex6plant}-\eqref{eq:ex6control}. From \eqref{eq:ex6coord1}, it is easy to see that for an initial condition \(\tilde\eta(0)=1.054\), no output $z$ that satisfies its limits can prevent $\tilde\eta$ from diverging. In \Cref{fig:exe_out}, this is demonstrated for a choice of $\alpha^{1}=(1,4,8,2)$ and $\bar x(0)=(-1.054,-0.072,5.483)$. Such a divergence occurs for all initial conditions with $|\tilde\eta(0)|>1$. It was also observed that for this choice of $\alpha^1$, the SIOCF condition was not met, with $\min_k\nu_k(\Omega)=-3.59$. The same phenomenon can be shown to occur for any choice of $\alpha^1$ given the internal dynamics in \eqref{eq:ex6coord1}. This underscores the importance of (A2) in \Cref{ass:standing}.

% Since (A2) of \Cref{ass:standing} is violated, \cref{alg:fix} is not applicable in this example. 

% The I-O CBF Governor parameters are therefore selected as $\alpha^{1}=(1,4,8,2)$. Under nominal input limits, corresponding to \(\varrho=1\), \Cref{fig:exe_out} illustrates that for the initial condition $\bar x(0)=(-1.054,-0.072,5.483)\in\Omega(1)$, the governor QP loses feasibility at \(t=0.120\). By relaxing QP constraints, the trajectory subsequently leaves \(\Omega(1)\) and the output eventually violates its limits. 

% The violation can be explained through the internal dynamics associated with the RHP invariant zero
% \begin{gather}
%     \dot{\tilde\eta}=\tilde\eta-z, \qquad
%     \tilde\eta=-x_1.
% \end{gather}
% For an initial condition \(\tilde\eta(0)=1.054\), no output $z$ that satisfies its limits can prevent $\tilde\eta$ from diverging.

% Let $\mathcal X_Z:=\{\xb:|\tilde\eta|\le1\}$ denote the set of initial conditions for which there exists a constraint-satisfying output $z$ that keeps the internal state $\eta$ bounded. A direct calculation yields $\Omega(\varrho)\subseteq\mathcal X_Z $ if and only if $\varrho\le\frac13$. 
% This threshold is also reflected in a numerical sweep of \(\min_k\nu_k(\Omega(\varrho))\), which yields nonnegative values for \(\varrho\le1/3\) and negative values for \(\varrho>1/3\).

\begin{figure}
    \centering
    \begin{subfigure}{1\columnwidth}
        \centering
        \includegraphics[width=\linewidth]{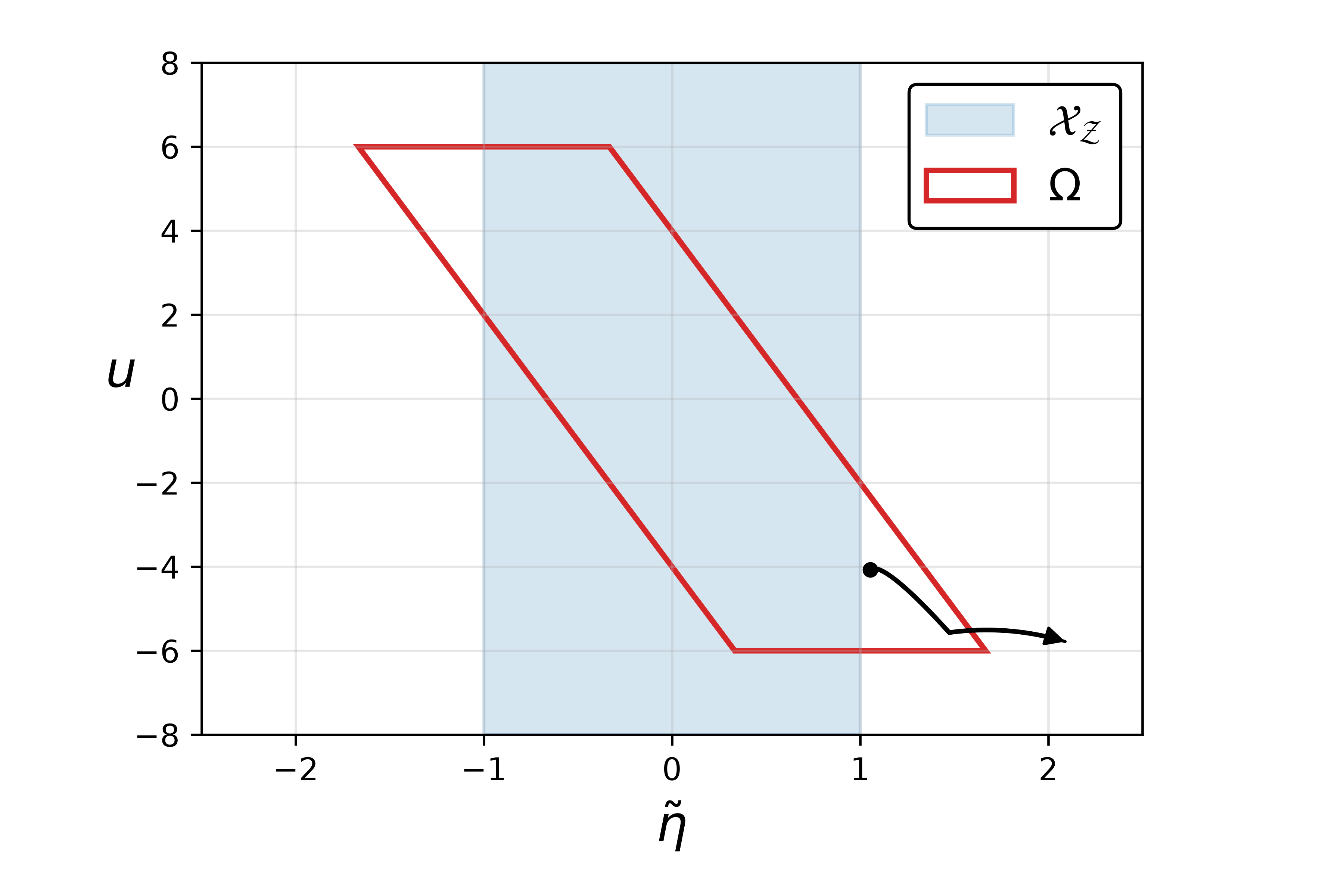}
        \vspace{-5mm}
    \end{subfigure}
    \begin{subfigure}{1\columnwidth}
        \centering
        \includegraphics[width=\linewidth]{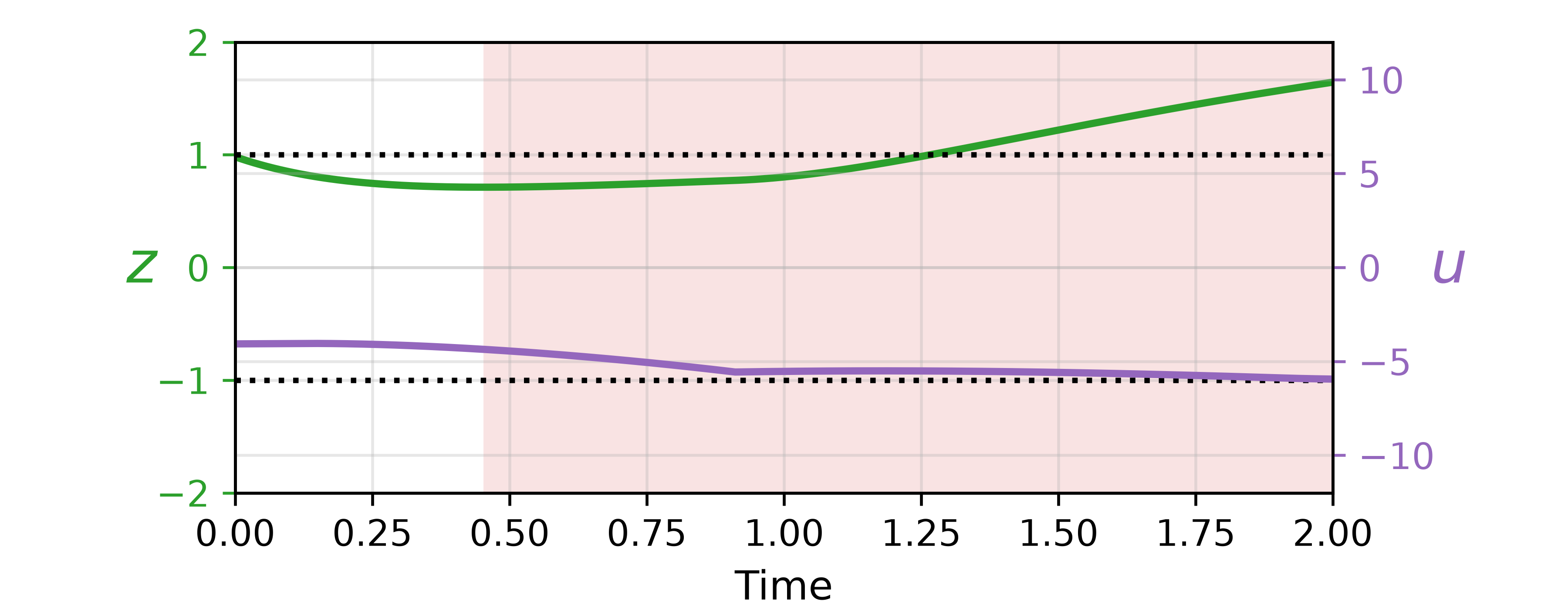}
    \end{subfigure}
    \caption{(Top) $\Omega$ is not completely contained within $\mathcal X_Z$, leading to initial conditions that can escape. (Bottom) Input and output constraints are not simultaneously satisfied.}
    \label{fig:exe_out}
\end{figure}

\endgroup

%===========================================%
%               CONCLUSIONS                 %
%===========================================%

\section{Conclusion}\label{sec:concl}
{This paper provides a numerical and simulation-based study of the feasibility framework developed in Part I \PartI\ for simultaneous input and output constraint satisfaction in a class of linear time-invariant systems with state feedback and integral action. The framework employs a Control Barrier Function (CBF) based governor that modifies the commanded reference while preserving the nominal feedback controller. Through a series of numerical examples, we demonstrate how the feasibility conditions derived in Part I can be evaluated offline and used to certify whether the governor remains feasible over a prescribed operating set. The examples include multi-input flight-control systems, open-loop unstable dynamics, and cases in which the natural operating set fails to be forward invariant. We further demonstrate cases in which simultaneous input and output constraints are fundamentally incompatible under the original actuator limits, and show how the constructive design procedure of Part I recovers feasibility through systematic selection of the CBF parameters and, when necessary, relaxation of the input constraints. The studies in this paper illustrate the necessity of the principal assumptions and conditions developed in Part I and demonstrate their use for both feasibility certification and governor design.}

\bibliographystyle{IEEEtran}
\bibliography{refs}

\clearpage
\onecolumn

\end{document}